\documentclass[preprint,showpacs,preprintnumbers,amsmath,amssymb]{revtex4}
\usepackage{graphicx}

\begin{document}
\def \ee {\varepsilon}
\thispagestyle{empty}
\title{Advance and prospects in constraining the Yukawa-type
corrections to Newtonian gravity from the Casimir effect}

\author{V.~B.~Bezerra,${}^{1}$
G.~L.~Klimchitskaya,${}^{2}$
V.~M.~Mostepanenko,${}^{3}$
and C.~Romero${}^{1}$
}

\affiliation{
${}^{1}$Department of Physics, Federal University of Para\'{\i}ba,
C.P.5008, CEP 58059--970, Jo\~{a}o Pessoa, Pb-Brazil \\
${}^{2}${North-West Technical University,
Millionnaya St. 5, St.Petersburg,
191065, Russia}\\
${}^3${Noncommercial Partnership ``Scientific Instruments'',
Tverskaya St. 11, Moscow,
103905, Russia}
}
\begin{abstract}
We report stronger constraints on the parameters of Yukawa-type
corrections to Newtonian gravity from measurements of the lateral
Casimir force between sinusoidally corrugated surfaces of a sphere
and a plate. In the interaction range from 1.6 to 14\,nm the
strengthening of previously known high confidence constraints
up to a factor of $2.4\times 10^7$ is achieved using these
measurements. It is shown that the replacement of a plane plate
with a corrugated one in the measurements of the normal Casimir
force by means of an atomic force microscope would result in the
strengthening of respective high confidence constraints on the
Yukawa-type interaction by a factor  of $1.1\times 10^{12}$.
The use of a corrugated plate instead of a plane plate in the
experiment by means of a micromachined oscillator also leads to
 strengthening of the obtained constraints. We further obtain
constraints on the parameters of Yukawa-type interaction from the
data of experiments measuring the gradient of the Casimir pressure
between two parallel plates and the gradient of the Casimir-Polder
force between an atom and a plate. The obtained results are
compared with the previously known constraints. The possibilities of
how to further strengthen the constraints on non-Newtonian
gravity are discussed.
\end{abstract}
\maketitle

\section{Introduction}

During the last few decades possible existence of
Yukawa-type corrections to Newtonian gravitational law
has attracted considerable attention \cite{1}.
In the middle of eighties the problem of the so-called
{\it fifth force} was widely discussed. Finally, no
deviations from the predictions of Newtonian gravity
have been found. However, after some period of time interest
in hypothetical corrections to Newton's law at short
separations was rekindled by numerous predictions of
high energy physics beyond the Standard Model. On the
one hand, many unification schemes predicted the
existence of massless and light bosons such as
arion \cite{2}, scalar axion \cite{3}, graviphoton \cite{4},
dilaton \cite{5}, goldstino \cite{6}, moduli \cite{7}, etc.
The exchange of light bosons between two atoms belonging
to different macrobodies generates an effective Yukawa-type
force at short range depending on the mass of the particle.
On the other hand, the Yukawa-type correction to Newton's
law at submillimeter separations was predicted by
multidimensional unification
 schemes, where additional spatial dimensions are
compactified at relatively low energy of the order of
1\,TeV \cite{8,9,10}. In such schemes, at separations much
larger than the size of a compact manifold, the gravitational
potential is given by the sum of Newton's and Yukawa-type
terms \cite{11,12}. Keeping in mind that at separations below
$10\,\mu$m the Newton's law was poorly tested experimentally,
the above theoretical predictions generated considerable
public excitement.

The key question is whether the Yukawa-type corrections to
Newtonian gravity exist in Nature and, if they do,  how strong they
are and what is their interaction range. The strongest
constraints on the parameters of these corrections in the
interaction range larger than $4.7\,\mu$m were obtained from
the E\"{o}tvos \cite{13,14} and Cavendish-type \cite{15,16,17,18}
 experiments. At shorter interaction range, however, the
gravitational experiments do not lead to competitive constraints
because the gravitational force becomes too weak.
As to very short separations, the hypothetical interactions of
Yukawa-type should be considered in a background of the
van der Waals and Casimir forces.

The possibility of obtaining constraints on the predicted forces
of Yukawa and power-type from the measurements of the Casimir
force was predicted in Refs.~\cite{19,20}, respectively.
The Casimir force originates from zero-point and thermal
fluctuations of the electromagnetic field. It acts at
separations of the order of $1\,\mu$m between the surfaces of
uncharged material bodies (see Ref.~\cite{21} for a recent
overview on the subject). The modern stage in the measurement
of the Casimir force has begun in 1997 and  has resulted in more
than twenty experiments (see review \cite{22}).
Many of them were used to obtain constraints on the
parameters of Yukawa-type interactions in the interaction
range from a few nanometers to a few micrometers.
In the torsion pendulum experiment \cite{23}
the constraints were obtained in Refs.~\cite{24,25};
in the experiments using an atomic force microscope
\cite{26,27,28} the respective constraints were found in
Refs.~\cite{29,30,31}, and in the experiment with two
crossed cylinders \cite{32} in Ref.~\cite{33}.
In all these experiments the Casimir force acting in
normal direction to the surfaces of a sphere and a plate
or two cylinders has been measured. As to the three
dynamic experiments using a micromachined oscillator
\cite{34,35,36},  the gradient of the normal Casimir
force acting between a sphere and a plate was measured.
In the proximity force approximation (PFA) this
gradient is proportional to the Casimir pressure in the
configuration of two parallel plates. Because of this,
the respective constraints were obtained on the Yukawa-type
pressure, rather than on the Yukawa-type force \cite{34,35,36}.

As is evident from the foregoing, the strongest constraints
on the parameters of Yukawa-type corrections to Newtonian
gravity at separations above a few micrometers are obtained
from gravitational experiments. Within a wide interaction
range from a few nanometers to a few micrometers, the
strongest constraints follow from the measurement of the
Casimir force. Notice that the first constraints on the
Yukawa-type hypothetical interaction obtained from the
Casimir effect \cite{19,24,25,29,30,31,33,34} were not
as exact and reliable as the constraints obtained from the
gravitational experiments at larger separations
\cite{13,14,15,16,17,18}. Specifically, for constraints
of Casimir origin the confidence levels were not determined.
This is due to some difficulties in the comparison
between experiment and theory when the measured force is
a strongly nonlinear function of separation.
Later, however, the use of appropriate statistical methods
\cite{21,22,35} allowed to obtain from the Casimir effect
the constraints of the same degree of reliability \cite{35,36}
as from the gravitational experiments. In addition, the
previously performed measurement of the Casimir force \cite{28}
was reanalyzed \cite{37} and respective constraints valid at
a 95\% confidence level were obtained \cite{36}. They are
slightly weaker than those in Ref.~\cite{31}, but benefit from
high confidence. It is pertinent to note that a widely debated
topic on the thermal contribution to the Casimir force \cite{21,22}
is irrelevant to constraining the hypothetical forces of
Yukawa-type from the Casimir effect because the difference
between the alternative thermal corrections considered in the
literature cannot be modeled by the Yukawa potential.
As a result, the measurements of the Casimir force have helped
to strengthen the previously known constraints on the Yukawa
interaction in the submicrometer range up to ten thousand
times.

In this paper we obtain stronger constraints on the
hypothetical interaction of Yukawa-type from recent measurement
of the lateral Casimir force between sinusoidally corrugated
surfaces of a sphere and a plate \cite{38}. As compared with
the previously known constraints in the interaction range below
14\,nm obtained at the same high confidence (a 95\% confidence
level), a strengthening up to a factor of
$2.4\times 10^7$ is achieved.
We also discuss respective limits on the parameters of hypothetical light
elementary particles. The use of corrugated surfaces opens new
prospective opportunities for constraining the Yukawa-type
hypothetical interactions from the Casimir effect. In view of this fact
in the present paper we propose several experiments of this
type and calculate the strength of constraints that can be
obtained in future in different interaction ranges.
Specifically we consider the measurement of the normal Casimir
force acting between a smooth sphere and a sinusoidally
corrugated plate. We also explore the potentials of dynamic
experiments where the separation distance between the test bodies
is varied harmonically and the measured quantities are the
gradients of the Casimir  force \cite{36} (Casimir-Polder
force \cite{39}),  or
of the Casimir pressure \cite{40}.

The structure of the paper is as follows. In Sec.~II we calculate
the lateral Yukawa force acting between sinusoidally corrugated
surfaces of a sphere and a plate and obtain constraints on its
parameters from the measurement results of Ref.~\cite{38}.
Section~III is devoted to the calculation of the normal Yukawa
force in the configuration of a smooth sphere above a sinusoidally
corrugated plate. Prospective constraints on the parameters of
this force obtainable in such a configuration are presented.
In Sec.~IV the gradient of the Yukawa force acting between a
smooth sphere and a corrugated plate in the dynamic regime is
calculated and used to estimate prospective constraints.
The possibility of using the gradient of the Yukawa pressure
between two parallel plates in the dynamic regime for obtaining
stronger constraints is considered in Sec.~V. In Sec.~VI
the gradient of the Yukawa force in the configuration of an
atom oscillating near a substrate is
calculated and applied for constraining
hypothetical interactions from the measurement data for the
Casimir-Polder force. In Sec.~VII the reader will find our
conclusions and discussions.

\section{Stronger constraints on the Yukawa-type hypothetical
{\protect \\}
interaction from the measurement of the lateral
Casimir force between corrugated surfaces}

It is customary to normalize the Yukawa interaction potential
between two neutral point masses $m_1$ and $m_2$ (atoms) at a
separation $r$ to the potential of Newtonian gravity and represent
it in the form \cite{21,41}
\begin{equation}
\frac{V^{\rm Yu}(r)}{V^{\rm N}(r)}=\alpha\,
e^{-r/\lambda}, \qquad
{V^{\rm N}(r)}=-\frac{Gm_1m_2}{r}.
\label{eq1}
\end{equation}
\noindent
Here, $G$ is the Newtonian gravitational constant, $\alpha$ is
a dimensionless constant characterizing the strength of the Yukawa
interaction and $\lambda$ is its interaction range.
Specifically, if the effective Yukawa potential between atoms
$m_1$ and $m_2$ is generated by the exchange of light bosons of
mass $m$, one has $\lambda=\hbar/(mc)$.

The lateral Casimir force acting between the sinusoidally
corrugated surfaces of a sphere and a plate was first
experimentally
demonstrated in Refs.~\cite{42,43}. In Ref.~\cite{43} a
measure of agreement between the data and the
theoretically calculated lateral Casimir force was used to
obtain constraints on the parameters of Yukawa interaction,
namely,
$\alpha$ and $\lambda$. This, however, did not lead to definitive
constraints determined at a high confidence level because
the determination
 of the lateral Casimir force between corrugated surfaces
was not sufficiently precise.

In a recent experiment \cite{38} the lateral Casimir force was
measured between two aligned sinusoidally corrugated surfaces of
a grating and a sphere with equal corrugation periods
$\Lambda=574.7\,$nm and corrugation amplitudes $A_1=85.4\,$nm
and $A_2=13.7\,$nm, respectively. The grating was made of  hard
epoxy with density $\rho_g=1.08\times 10^3\,\mbox{kg/m}^3$
on a 3\,mm thick Pyrex substrate. The top of the grating was
covered with a $\Delta_{{\rm Au},g}=300\,$nm thick Au coating of
 density $\rho_{\rm Au}=19.28\times 10^3\,\mbox{kg/m}^3$.
 The sphere of $R=97.0\,\mu$m radius was made of polystyrene
 of  density $\rho_s=1.06 \times 10^3\,\mbox{kg/m}^3$ and
 uniformly coated with a $\Delta_{\rm Cr}=10\,$nm layer of Cr
 ($\rho_{Cr}=7.14\times 10^3\,\mbox{kg/m}^3$) and then with
a $\Delta_{{\rm Au},s}=50\,$nm layer of Au in a thermal
evaporator \cite{38}.

The lateral Casimir force was independently (without fit to any
theory) measured as a function of the phase shift $\varphi$
between corrugations on the sphere and on the grating over the
region of separations, $a$,
in the range  from 120 to 190\,nm. At each separation
$a_i$ the maximum magnitude of the lateral Casimir force was
achieved at some phase shift $\varphi_i$ (keeping in mind that
the dependence of the lateral force on $\varphi$ is not strictly
sinusoidal, $\varphi_i$ must not be  multiples of $\pi/2$).
The absolute errors, $\Delta_i$,
of the measured maximum magnitudes of the lateral
Casimir force at different separations were determined
at a 95\% confidence level. The measurement data were compared
with the exact theory taking into account a nonzero skin depth of
the Au coating and treating the Casimir force due to the nontrivial
geometry of sinusoidally corrugated surfaces in the framework
of the Rayleigh scattering approach. It was found that the
experimental data  agree with the theory within the limits of the
experimental errors $\Delta_i$.
This means that the magnitudes of any possible lateral Yukawa-type
force that might arise between the corrugated surfaces of a sphere
and a grating must satisfy the inequality
\begin{equation}
\left|F_{ps,\rm cor}^{\rm Yu,lat}(a_i,\varphi_i)\right|
\leq\Delta_i,
\label{eq2}
\end{equation}
\noindent
with $\Delta_i$ as defined above. Equation (\ref{eq2}) follows
from the fact that the measured force consisting of the lateral
Casimir force and possible lateral Yukawa-type force agrees
with the theoretical lateral Casimir force to within the
experimental error of the force measurement (note that in the
experiment under consideration the experimental errors $\Delta_i$
are much larger than the errors in computations of the lateral
Casimir force with the help of the scattering approach).

To obtain constraints on the parameters of the Yukawa-type
interaction $\alpha$ and $\lambda$, following from
Eq.~(\ref{eq2}),
we should substitute in this
equation an explicit expression for the
lateral Yukawa-type force. This can be found in the following
way. We first consider a smooth homogeneous sphere of density
$\rho_s$  placed
at a distance $a$ (in the $z$ direction) above a plane
plate of density $\rho_g$ with thickness $D$ and linear
dimension $L$. Under the conditions $a,\,\lambda\ll D,\,L,\,R$
which are
satisfied in our case with large supply, one can consider a plate
of  infinite area and thickness (i.e., a semispace).
The expression for the Yukawa-type interaction energy and force
between a thick plate  (semispace)
and a sphere spaced at a separation $a$ are
obtained by the integration of $V^{\rm Yu}(r)$ in Eq.~(\ref{eq1})
over both volumes and subsequently
taking the  negative differentiation with
respect to $a$ \cite{30}:
\begin{eqnarray}
E_{ps}^{\rm Yu}(a)&=&-4\pi^2G\alpha\lambda^4\rho_g\rho_s
e^{-a/\lambda}\Phi(R,\lambda),
\nonumber \\
F_{ps}^{\rm Yu}(a)&=&-4\pi^2G\alpha\lambda^3\rho_g\rho_s
e^{-a/\lambda}\Phi(R,\lambda),
\label{eq3}
\end{eqnarray}
\noindent
where
\begin{equation}
\Phi(x,\lambda)\equiv x-\lambda+(x+\lambda)e^{-2x/\lambda}.
\label{eq4}
\end{equation}
\noindent
Applying the first equality in (\ref{eq3}) to the layer
structure of the experimental test bodies, as described above,
one arrives at
\begin{eqnarray}
&&
E_{ps,l}^{\rm Yu}(a)=-4\pi^2G\alpha\lambda^4
e^{-a/\lambda}\Psi(\lambda),
\label{eq5} \\
&&
\Psi(\lambda)\equiv\left[\rho_{\rm Au}-(\rho_{\rm Au}-\rho_g)
e^{-\Delta_{{\rm Au},g}/\lambda}\right]
\nonumber \\
&&
~~\times\left[\rho_{\rm Au}\Phi(R,\lambda)-
(\rho_{\rm Au}-\rho_{\rm Cr})\Phi(R-\Delta_{{\rm Au},s},\lambda)
e^{-\Delta_{{\rm Au},s}/\lambda}\right.
\nonumber \\
&&
~~~~\left.
-(\rho_{\rm Cr}-\rho_{s})
\Phi(R-\Delta_{{\rm Au},s}-\Delta_{\rm Cr},\lambda)
e^{-(\Delta_{{\rm Au},s}+\Delta_{\rm Cr})/\lambda}\right].
\nonumber
\end{eqnarray}
\noindent
As shown below, significant strengthening of constraints on
$\alpha$ from the data of this experiment holds at $\lambda$
below 10\,nm. Because of this, keeping in mind that
$\Delta_{{\rm Au},g}=300\,$nm, not only the thickness of Pyrex
substrate but the Au layer on the grating as well can be considered
as infinitely thick.

Now we need to calculate the effect of aligned corrugations on
the Yukawa-type energy (\ref{eq5}). This can be done by using the
method of geometrical averaging \cite{21,22}, i.e., by replacing
the closest separation $a$ between the smooth surfaces in
Eq.~(\ref{eq5}) with respective separation between corrugated
surfaces and averaging over the period of corrugations.
In order for such an approximate method  works properly,
in addition to the conditions indicated above, one more condition,
namely,
$\Lambda\ll R$, should be valid. It is satisfied with large
supply for the listed above
experimental parameters of Ref.~\cite{38}.

The separation distance between the closest points of corrugated
surfaces is given by
\begin{equation}
z_2-z_1=a+A_2\sin(2\pi x/\Lambda+\varphi)-A_1\sin(2\pi x/\Lambda).
\label{eq6}
\end{equation}
\noindent
It can be identically represented as
\begin{equation}
z_2-z_1=a+b\cos(2\pi x/\lambda-\tilde{\varphi}),
\label{eq7}
\end{equation}
\noindent
where the following notations are introduced:
\begin{eqnarray}
&&
b\equiv b(\varphi)=(A_1^2+A_2^2-2A_1A_2\cos\varphi)^{1/2},
\label{eq8} \\
&&
\tan\tilde{\varphi}=(A_2\cos\varphi-A_1)/(A_1\sin\varphi).
\nonumber
\end{eqnarray}
\noindent
Substituting Eq.~(\ref{eq7}) into Eq.~(\ref{eq5}) in place of $a$
and performing the geometrical averaging, we arrive at
\begin{equation}
E_{ps,\rm cor}^{\rm Yu}(a,\varphi)=
-4\pi^2G\alpha\lambda^4\Psi(\lambda)
\frac{e^{-a/\lambda}}{\Lambda}\int_{0}^{\Lambda}dx\,
e^{-b\cos(2\pi x/\Lambda-\tilde\varphi)/\lambda}.
\label{eq9}
\end{equation}
\noindent
The integral in Eq.~(\ref{eq9}) can be calculated using the formula
2.5.40(3) in Ref.~\cite{44}, with the result
\begin{equation}
E_{ps,\rm cor}^{\rm Yu}(a,\varphi)=
-4\pi^2G\alpha\lambda^4\Psi(\lambda)
\,{e^{-a/\lambda}}{I}_0(b/\lambda),
\label{eq10}
\end{equation}
\noindent
where ${I}_n(z)$ is the Bessel function of imaginary
argument. The lateral Yukawa force between corrugated surfaces
is obtained from Eq.~(\ref{eq10}) by
taking the negative differentiation
with respect to the phase shift:
\begin{eqnarray}
&&
F_{ps,\rm cor}^{\rm Yu,lat}(a,\varphi)=-\frac{2\pi}{\Lambda}\,
\frac{\partial E_{ps,\rm cor}^{\rm Yu}(a,\varphi)}{\partial\varphi}
\nonumber \\
&&~~~
=8\pi^3G\alpha\lambda^3\Psi(\lambda)
\,{e^{-a/\lambda}}\frac{A_1A_2}{b\Lambda}\,
{I}_1(b/\lambda)\sin\varphi.
\label{eq11}
\end{eqnarray}

Now we are  in a position to determine the constraints on the
parameters
of the Yukawa-type interaction  following from
the measurement data of experiment \cite{38}. For this purpose
we substitute Eq.~(\ref{eq11}) into Eq.~(\ref{eq2}) and find
the allowed values of $\alpha$ and $\lambda$.
The computational results are shown in Fig.~1, where the allowed
values of $\alpha,\,\,\lambda$ lie below the solid line and the
prohibited values of $\alpha,\,\,\lambda$ lie above the solid
line. These constraints were obtained in the interaction
region ranging from $\lambda=1.6\,$nm to $\lambda=35.5\,$nm using the
measurement data of Ref.~\cite{38} at different separations $a_i$.
Thus, at $\lambda\leq 4.5\,$nm the strongest constraints follow
from Eq.~(\ref{eq2}) with $a_1=121.1\,$nm and $\Delta_1=11.1\,$pN.
For $\lambda$ in the range
from 4.5 to 22.4\,nm, the separation $a_2=124.7\,$nm
has been used with $\Delta_2=4.7\,$pN. Finally, for
$\lambda$ from 22.4 to 35.5\,nm the strongest constraints were
obtained for $a_3=137.3\,$nm and $\Delta_3=2.5\,$pN \cite{38}.

Note that our constraints indicated by the solid line are
determined with the same 95\% confidence level as the absolute
errors $\Delta_i$ on the right-hand side
of Eq.~(\ref{eq2}). For comparison purposes in the same figure
we have plotted the previously known strongest constraints on
$\alpha,\,\,\lambda$ determined with 95\% confidence level.
The long-dashed line follows from the measurement data of
the experiment \cite{28} reanalyzed using rigorous statistical
methods in Refs.~\cite{36,37}. The short-dashed line represents
constraints obtained from the experiment \cite{36}.
As can be seen in Fig.~1, our constraints indicated by the solid
line are the strongest over the interaction range from 1.6 to
14\,nm. The largest strengthening of previously known constraints
shown by the long-dashed line by a factor of $2.4\times 10^7$ is
achieved at $\lambda=1.6\,$nm. (Note that the confidence level
of slightly stronger constraints obtained in Ref.~\cite{33}
from the measurement of the Casimir force between two crossed
cylinders \cite{32} cannot be determined because of several
uncertainties inherent to this experiment \cite{21,22}.)
The physical reason for so strong strengthening
of the constraints obtained from the
experiment with corrugated surfaces is that at a separation, for
instance, $a=121.4\,$nm between the mean levels of corrugations,
the distance between two closest points of the surfaces can
be as small as only 22\,nm.

The strongest constraints on the Yukawa-type interaction
shown in Fig.~1 place limits on the parameters of gauge
barions and strange moduli \cite{7,45}. The existence
of such particles, which are predicted in many
extra-dimensional models, would result in a Yukawa-type
interaction with a very large $|\alpha|$ and an interaction range
$\lambda$ from $10^{-8}\,$m to $3\times 10^{-6}\,$m.
The obtained results can be also used to constrain the
predictions of chameleon theories which introduce scalar
fields with masses depending on the local background matter
density \cite{46,47}.

\section{Yukawa force between a smooth sphere and
{\protect \\}
a corrugated plate}

The constraints on $\alpha,\,\,\lambda$ shown by the long-dashed
line in Fig.~1 are obtained from the measurement of the normal
Casimir force (i.e., directed perpendicular to the surface) acting
between a smooth sphere and a plane plate \cite{28}.
The question arises whether the use of a corrugated plate
(a grating) could lead to stronger constraints. The normal
Casimir forces acting between a smooth sphere and a corrugated
plate with sinusoidal \cite{48} and rectangular \cite{49}
corrugations have been measured. However, due to the absence of
sufficiently exact theory of the Casimir force
applicable to corrugated surfaces
 at that time, it was not possible to
investigate the strength of constraints which might be imposed
using the results of these measurements. Presently the exact
theory of the Casimir force applicable to corrugated surfaces
of test bodies made of real materials, at the laboratory
temperature, is available \cite{38}. Because of this, it is
pertinent to verify the potentialities of different
configurations with corrugated surfaces for the strengthening of
constraints on the Yukawa-type interaction. In this section we
consider the configuration of a smooth sphere above a
sinusoidally corrugated plate (grating).

The energy of the Yukawa-type interaction for the configuration of
our interest is obtainable from Eq.~(\ref{eq10}), where the
amplitude of corrugations on a sphere is put equal to zero,
$A_2=0$, and from Eq.~(\ref{eq8}) it follows that $b=A_1$.
As a result
\begin{equation}
E_{p,\rm cor}^{\rm Yu}(a)=
-4\pi^2G\alpha\lambda^4\Psi(\lambda)
\,{e^{-a/\lambda}}{I}_0(A_1/\lambda).
\label{eq12}
\end{equation}
\noindent
 The respective Yukawa-type force is given by
\begin{eqnarray}
&&
F_{p,\rm cor}^{\rm Yu}(a)=-
\frac{\partial E_{p,\rm cor}^{\rm Yu}(a)}{\partial a}
\nonumber \\
&&~~~
=-4\pi^2G\alpha\lambda^3\Psi(\lambda)
\,{e^{-a/\lambda}}
{I}_0(A_1/\lambda).
\label{eq13}
\end{eqnarray}
\noindent
The constraints on $\alpha,\,\,\lambda$ can be found from the
inequality
\begin{equation}
\left|F_{p,\rm cor}^{\rm Yu}(a)\right|\leq\Xi_{F}(a).
\label{eq14}
\end{equation}
\noindent
Here, the confidence interval $[-\Xi_F(a),\Xi_F(a)]$ for the
difference between theoretical and mean experimental Casimir
forces, $F^{\rm theor}(a)-\bar{F}^{\rm expt}(a)$, was found
at different separations in Ref.~\cite{37} at a 95\%
confidence level. As an example, at separations $a=100$,
110 and 120\,nm, the half-width of the confidence interval
is equal to 9.17, 8.35, and 7.96\,pN, respectively.

The computations of the constraints were performed by the
substitution of Eq.~(\ref{eq13}) into Eq.~(\ref{eq14}) for the
 realistic values of the  parameters given below. The corrugated plate
(grating) was precisely the same as described in Sec.~II basing
 on the experimental configuration of Ref.~\cite{38}.
The smooth polystyrene sphere of radius $R=100\,\mu$m was
assumed to be coated with only one layer of Au (as in the
experiment of Ref.~\cite{28}) of thickness
$\Delta_{{\rm Au},s}=100\,$nm. This means that in the expression
(\ref{eq5}) for the function $\Psi(\lambda)$ one must put
$\Delta_{\rm Cr}=0$. The obtained constraints are shown by the
grey line in Fig.~2 (as in Fig.~1, the region of
$\alpha,\,\,\lambda$ below the line is allowed and above the
line is prohibited). For comparison purposes, we reproduce in
Fig.~2 the solid and the short-dashed lines of Fig.~1, which
show the constraints obtained by us in Sec.~II from the
measurement of the lateral Casimir force and in Ref.~\cite{36}
from the experiment using a micromachined oscillator,
respectively. As is seen in Fig.~2, the largest strengthening of
already obtained constraints given by the solid line up to a
factor of $4.5\times 10^4$ occurs at $\lambda=1.6\,$nm.
It is obtained from the measurement data at the shortest
separation $a=100\,$nm.

It is instructive also to compare the prospective constraints
given by the grey line in Fig.~2 with those obtained from
the experiment \cite{28} in Ref.~\cite{36} using the
configuration of a smooth sphere above a plane plate (the
long-dashed line in Fig.~1). Here, at $\lambda=1.6\,$nm the
largest strengthening of constraints by a factor of
$1.1\times 10^{12}$ is possible. So strong improvement
of the obtained constraints can be achieved due to only
a replacement of the plane plate with the sinusoidally
corrugated plate.
The physical reason for so large strengthening is the same as
in Sec.~II: when the plate is corrugated the closest
separation between the two surfaces is as small as 14.6\,nm.
This demonstrates that the use of
corrugated test bodies is of high promise for obtaining
stronger constraints on the Yukawa-type hypothetical
interactions from the measurements of the Casimir force.

\section{Gradient of the Yukawa force between a smooth sphere and
a corrugated plate in the dynamic regime}

The most precise experiment in Casimir physics was performed in
the dynamic regime by means of micromechanical torsional
oscillator \cite{36}. This experiment exploited the configuration
of a smooth sphere of $R=150\,\mu$m radius above a plane plate
which could rotate about a torsional axis. The separation
distance between them was varied harmonically with the
resonant frequency of the oscillator.
This frequency is different in the absence and in the presence
of the Casimir force $F_{sp}(a)$ between a sphere and a plate.
The immediately measured quantity is the frequency shift due to
the presence of the sphere which is proportional to the gradient
of $F_{sp}(a)$. Using the proximity force approximation (PFA),
which is valid in the experimental configuration with an
accuracy of about 0.1\% \cite{21,22}, one obtains
\begin{equation}
P(a)=-\frac{1}{2\pi R}\,
\frac{\partial F_{sp}(a)}{\partial a},
\label{eq15}
\end{equation}
\noindent
where $P(a)$ is the Casimir pressure in the fictitious
configuration of two plates having the same layer structure as the
plate and the sphere in real experiment. In so doing, the
lower plate simply coincides with the real plate. It can be
replaced with a semispace if it is sufficiently thick. The upper
(fictitious) plate is necessarily infinitely thick, i.e., is
always a semispace. Thus, the experiment of Ref.~\cite{36}
can be considered as an indirect measurement of the Casimir
pressure in the configuration of two parallel plates.

The constraints on the parameters of Yukawa-type interaction
in Ref.~\cite{36} were obtained from the condition
\begin{equation}
\left|P^{\rm Yu}(a)\right|\leq\Xi_P(a),
\label{eq16}
\end{equation}
\noindent
where $P^{\rm Yu}(a)$ is the Yukawa pressure in the configuration
of two parallel plates and $[-\Xi_P(a),\Xi_P(a)]$ is the
minimum confidence interval containing all differences between
theoretical and mean experimental pressures,
$P^{\rm theor}(a)-\bar{P}^{\rm expt}(a)$, within the
separation region from 180 to 746\,nm. This confidence
interval was determined at a 95\% confidence level.
The use of Eq.~(\ref{eq16}) for obtaining the constraints
assumes that the PFA is applicable to calculate not only the
Casimir force but the Yukawa-type force as well. This
applicability was
confirmed in Ref.~\cite{50}, where the Yukawa-type force in
the experimental configuration \cite{36} was calculated both
exactly and using the PFA with coinciding results up to a 0.1\%
error. Thus it was shown that the PFA leads to the results
of the same level of precision when applied to the Casimir
and Yukawa-type forces. However, in Ref.~\cite{51}
it is claimed that to
obtain correct constraints on the Yukawa-type force the latter must
be calculated at least ten times more precisely than the
Casimir force. This claim was demonstrated to be incorrect in
Ref.~\cite{52}, where the constraints were reobtained without
using the PFA. Instead, the inequality was used
\begin{equation}
\frac{1}{2\pi R}\,\left|
\frac{\partial F_{sp}^{\rm Yu}(a)}{\partial a}\right|\leq\Xi_P(a),
\label{eq17}
\end{equation}
\noindent
where $F_{sp}^{\rm Yu}(a)$ is the exact Yukawa force in a
sphere-plate configuration. The obtained constraints
were shown to coincide with
those found in Ref.~\cite{36} up to three significant figures.
They are represented as the solid line in Fig.~3
(note that these constraints were already represented as
short-dashed lines in Figs.~1 and 2).
We emphasize that for the purpose of constraining parameters of
the Yukawa-type force, it should never need to be calculated more
precisely than up to 0.1\% because the border of the confidence
interval is determined up to only two or, at maximum, three
significant figures.

Here, we check whether it is possible to strengthen the
constraints of Ref.~\cite{36} by replacing a plane plate of
the torsional oscillator with a corrugated plate. The parameters
of the latter are assumed to be the same as in Sec.~II.
We perform the calculation of the prospective constraints exactly,
i.e., we do not use the PFA. The Yukawa-type force acting between a
smooth sphere and a corrugated plate is given by Eq.~(\ref{eq13}).
Thus, for the gradient of the Yukawa force one obtains
\begin{equation}
\frac{\partial F_{sp}^{\rm Yu}(a)}{\partial a}=
4\pi^2G\alpha\lambda^2\Psi(\lambda)e^{-a/\lambda}
{I}_0(A_1/\lambda).
\label{eq18}
\end{equation}
\noindent
The function $\Psi(\lambda)$ is defined in Eq.~(\ref{eq5})
where in the experimental configuration of Ref.~\cite{36}
$\Delta_{{\rm Au},s}=180\,$nm and $\Delta_{\rm Cr}=10\,$nm.
Then, the expression for the force gradient (\ref{eq18})
was substituted into Eq.~(\ref{eq17}). The obtained
prospective constraints are shown in Fig.~3 as the grey
line. The best constraints in the interaction range
$\lambda\leq 80\,$nm were obtained from $\Xi_P=3.30\,$mPa
at $a=200\,$nm \cite{21,36}. For the interaction range
from 80 to 140\,nm and from 140 to 250\,nm the constraints
shown by the grey line were obtained from $\Xi_P=0.84$
and 0.57\,mPa, as occurs at $a=300$ and 350\,nm,
respectively. For comparison purposes in the same figure
the long-dashed line reproduces constraints following
from the experiment of Ref.~\cite{28} and reobtained
in Refs.~\cite{36,37} at a 95\% confidence level.
The short-dashed line shows constraints following from
the so-called {\it Casimir-less} experiment \cite{53}
where the contribution of the Casimir force is
subtracted due to the advantages of the dynamic
measurement scheme. As can be seen in Fig.~3, the
strengthening of constraints on the parameters of
Yukawa-type interaction due to the use of corrugated
plate is up to one order of magnitude. The largest
strengthening (in 10 times) is achieved at
$\lambda=18.6\,$nm. If we compare with the static
experiment proposed in Sec.~III, the use of a plate
covered with sinusoidal corrugations turns out to be
not so promising (one order of magnitude strengthening
of constraints instead of a factor of $4.5\times 10^4$
strengthening). This can be explained by the fact
that the measurement technique using a micromachined
oscillator is workable only in the region where this
oscillator is linear, i.e., starting from at least
two times larger separations than using an AFM.
Here, the shortest separation between a corrugated plate
and a smooth sphere is equal to 114.6\,nm (compare
with 22\,nm and 14.6\,nm in Secs.~II and III, respectively).

\section{Gradient of the Yukawa pressure between two parallel
plates in the dynamic regime}

The only recent experiment on measuring the Casimir force
which used the configuration of two parallel plates was
performed in the dynamic regime \cite{40}.
The immediately measured quantity was the oscillator frequency
shift due to the Casimir pressure between the plates. This
frequency shift is proportional to the gradient of the
Casimir pressure
\begin{equation}
\nu^2-\nu_0^2=-\beta\,\frac{\partial P(a)}{\partial a},
\label{eq19}
\end{equation}
\noindent
where the coefficient is equal to
$\beta\approx 0.0479\,\mbox{m}^2/$kg.
Until the present time the constraints on the Yukawa-type
interaction following from this experiment have not been
published. Because of this, here we derive these constraints
and discuss the possibilities of their strengthening.

We start with the expression for the Yukawa-type pressure
between two parallel plates made of Si of density
$\rho_{\rm Si}=2.33\times 10^3\,\mbox{kg/m}^3$ and covered
with chromium layers of thickness
$\Delta_{\rm Cr}=50\,$nm, as in the experiment of Ref.~\cite{40}.
Keeping in mind that the experimental parameters satisfy the
conditions $a,\,\lambda\ll D_i,\,L_i$, where $D_{1,2}$ and
$L_{1,2}$ are the thicknesses and linear dimensions of both
plates, one can consider these plates as semispaces.
The integration of the Yukawa-type potential (\ref{eq1})
over the volumes of both plates leads to \cite{21,50}
\begin{equation}
E^{\rm Yu}(a)=-2\pi G\alpha\lambda^3\rho_{\rm Si}^2
e^{-a/\lambda}.
\label{eq20}
\end{equation}
\noindent
Applying this equation to the plates covered with chromium
layers one obtains
\begin{equation}
E_l^{\rm Yu}(a)=-2\pi G\alpha\lambda^3e^{-a/\lambda}
\left[\rho_{\rm Cr}-(\rho_{\rm Cr}-\rho_{\rm Si})
e^{-\Delta_{\rm Cr}/\lambda}\right]^2.
\label{eq21}
\end{equation}
\noindent
This leads to the following expression for the Yukawa
pressure
\begin{eqnarray}
P_l^{\rm Yu}(a)&=&
-\frac{\partial E_l^{\rm Yu}(a)}{\partial a}
\nonumber \\
&=&-2\pi G\alpha\lambda^2e^{-a/\lambda}
\left[\rho_{\rm Cr}-(\rho_{\rm Cr}-\rho_{\rm Si})
e^{-\Delta_{\rm Cr}/\lambda}\right]^2
\label{eq22}
\end{eqnarray}
\noindent
and for the magnitude of its gradient
\begin{equation}
\left|\frac{\partial P_l^{\rm Yu}(a)}{\partial a}\right|
=2\pi G|\alpha|\lambda e^{-a/\lambda}
\left[\rho_{\rm Cr}-(\rho_{\rm Cr}-\rho_{\rm Si})
e^{-\Delta_{\rm Cr}/\lambda}\right]^2.
\label{eq23}
\end{equation}

The constraints on the Yukawa-type interaction can be obtained
from the inequality
\begin{equation}
\beta\left|\frac{\partial P_l^{\rm Yu}(a_i)}{\partial a}\right|
\leq\Delta_i(\nu^2-\nu_0^2),
\label{eq24}
\end{equation}
\noindent
where $\Delta_i(\nu^2-\nu_0^2)$ is the absolute error in the
measurement of the frequency shift $\nu^2-\nu_0^2$ determined
in Ref.~\cite{40} at different separations $a_i$ where
measurements were performed at a 67\% confidence level.
Now we substitute Eq.~(\ref{eq23}) into Eq.~(\ref{eq24})
and determine the allowed and prohibited regions of the
parameters $\alpha$ and $\lambda$. These regions are
separated by the solid line in Fig.~4. The best constraints
shown by the solid line were obtained using the experimental
data at different separation distances.
Thus, at $\lambda\leq 0.2\,\mu$m
the value $\Delta_1(\nu^2-\nu_0^2)=67.1\,\mbox{Hz}^2$ that was
used  occurs at $a_1=0.553\,\mu$m. For interaction ranges
from $\lambda=0.2$ to $0.35\mu$m and $\lambda\geq 0.35\,\mu$m
we have used the values
$\Delta_2(\nu^2-\nu_0^2)=58.4\,\mbox{Hz}^2$ and
$\Delta_3(\nu^2-\nu_0^2)=19.2\,\mbox{Hz}^2$,
 which occur at separations  $a_2=0.574\,\mu$m and
 $a_3=0.8805\,\mu$m, respectively \cite{40}.
 For comparison purposes, in Fig.~4 we present also the
 constraints on the Yukawa-type interaction obtained in
 Ref.~\cite{53} from the Casimir-less experiment (the short-dashed
 line) and the constraints obtained in Ref.~\cite{24} from the
 torsion pendulum experiment of Ref.~\cite{23} (the long-dashed
 line). The latter constraints will be discussed in more detail
 in the next section. As can be seen in Fig.~4, the constraints
 following from the measurement of the gradient of the Casimir
 pressure between two parallel plates are not as strong as the
 constraints obtained from previously performed experiments.

 The experimental configuration of two parallel plates has some
 potentialities to obtain stronger constraints on the parameters
 of the Yukawa-type interaction. One evident resource is
 connected with the increase of precision [i.e.,with the
 decrease of $\Delta_i(\nu^2-\nu_0^2)$]. Furthermore, the
 replacement of Cr, as the material of metal coating, with an Au
 layer of the same thickness would lead to stronger constraints
 shown in Fig.~4 by the grey line marked 1. Moreover, the increase
 of thicknesses of the metal coating up to $\Delta_{\rm Au}=500\,$nm
 would lead to even stronger constraints shown in Fig.~4 by the
 grey line marked 2. This line presents constraints of the same
 strength as those obtained in Ref.~\cite{53} from the
Casimir-less experiment. Thus, the configuration of two parallel
plates can be further used for obtaining constraints on the
Yukawa-type interaction. As to the possibility of replacing a
plane plate with a corrugated one, this does not lead to
significantly stronger constraints in the experiment with
parallel plates because of much larger separation distances
between the test bodies.

\section{Constraints from the measurement of thermal
{\protect \\}
Casimir-Polder force}

Measurements of the Casimir-Polder force between an atom and
a plate (cavity wall) and the measure of their agreement with
theory provides additional opportunities for constraining the
Yukawa-type corrections to Newtonian gravity. Up to now no
strongest constraints have been obtained in this way, but to a large
extent the potentialities of this field remain unexplored. Here,
we obtain constraints on the parameters of the Yukawa-type
interaction using the measurement data of dynamic experiment
demonstrating the thermal Casimir-Polder force between
${}^{87}$Rb atoms and a SiO${}_2$ plate \cite{39}.
Rubidium atoms belonged to a Bose-Einstein condensate which
was produced in a magnetic trap with frequencies equal to
$\omega_{0z}=1438.85\,$rad/s and $\omega_{0l}=40.21\,$rad/s
in the perpendicular and longitudinal directions to the plate,
respectively. This resulted in  Thomas-Fermi radii
$R_z=2.69\,\mu$m and $R_l=97.1\,\mu$m, respectively.
 An oscillation amplitude
$A_z=2.50\,\mu$m in the $z$ direction was chosen and was kept
constant. By illuminating the plate with laser pulses it was
possible to vary its temperature. During measurements of the
Casimir-Polder force, the separation distance between the cloud
of ${}^{87}$Rb atoms and the plate was varied in the range from
7 to $11\,\mu$m. Due to the influence of the Casimir-Polder
force, the oscillation frequency in the $z$ direction
shifted, and the relative frequency shift
\begin{equation}
\gamma_z=\frac{|\omega_{0z}-\omega_z|}{\omega_{0z}}\approx
\frac{|\omega_{0z}^2-\omega_z^2|}{2\omega_{0z}^2}
\label{eq24a}
\end{equation}
\noindent
was measured as the function of the separation $a$ between the plate
and the center of mass of the condensate. The absolute errors in
the measurement of $\gamma_z$ at different separations $a_i$,
$\Delta_i\gamma_z$, were found in Ref.~\cite{39} at a 67\%
confidence level.

The Yukawa-type interaction (if any) would also lead to some
frequency shift in the perpendicular direction to the plate.
It can be calculated in the following way. The Yukawa energy of
a single atom of mass $m_1$ above a thick plate can be found
by the integration of the potential (\ref{eq1}) over the
volume of the plate
\begin{equation}
E_{ap}^{\rm Yu}(a)=-2\pi G\alpha\lambda^2m_1\rho_2e^{-a/\lambda},
\label{eq25}
\end{equation}
\noindent
where $\rho_2=2.203\times 10^3\,\mbox{kg/m}^3$ is the density of
plate material (fused silica). {}From this the Yukawa-type force
acting on an atom is given by
\begin{equation}
F_{ap}^{\rm Yu}(a)=
-\frac{\partial E_{ap}^{\rm Yu}(a)}{\partial a}
=-2\pi G\alpha\lambda m_1\rho_2e^{-a/\lambda}.
\label{eq26}
\end{equation}
\noindent
The frequency shift due to the Yukawa force (\ref{eq26}) can be
calculated by averaging over the deviations of separate
atoms  in the
$z$-direction from the center of mass of a condensate, which
is taken at the origin of the $z$ axis, and over the oscillation
period \cite{21,54,55}
\begin{eqnarray}
&&
\omega_{0z}^2-\omega_z^2=-\frac{\omega_{0z}}{\pi A_zm_1}
\int_{0}^{2\pi/\omega_{0z}}d\tau\,\cos(\omega_{0z}\tau)
\nonumber \\
&&~~~\times
\int_{-R_z}^{R_z}dz\,n_z(z)
F_{ap}^{\rm Yu}[a+z+A_z\cos(\omega_{0z}\tau)].
\label{eq27}
\end{eqnarray}
\noindent
Here, the distribution function of the atomic gas density
is given by
\begin{equation}
n_z(z)=\frac{15}{16R_z}\left(1-\frac{z^2}{R_z^2}\right)^2.
\label{eq28}
\end{equation}
\noindent
Substituting Eq.~(\ref{eq26}) into Eq.~(\ref{eq27}) we can find
the relative frequency shift due to the Yukawa-type interaction
defined in the same way as in Eq.~(\ref{eq24a})
\begin{eqnarray}
\gamma_z^{\rm Yu}(a)&=&\frac{G\alpha\lambda\rho_2}{\omega_{0z}A_z}
\,\frac{15}{16R_z}\,e^{-a/\lambda}
\int_{-R_z}^{R_z}dz\,\left(1-\frac{z^2}{R_z^2}\right)^2
e^{-z/\lambda}
\nonumber \\
&\times &
\int_{0}^{2\pi/\omega_{0z}}d\tau\,\cos(\omega_{0z}\tau)
e^{-A_z\cos(\omega_{0z}\tau)/\lambda}.
\label{eq29}
\end{eqnarray}
\noindent
Calculating both integrals on the right-hand side of
Eq.~(\ref{eq29}), one obtains \cite{44}
\begin{equation}
\gamma_z^{\rm Yu}(a)=
\frac{15\pi G\lambda\rho_2}{8\omega_{0z}^2A_z}
\alpha e^{-a/\lambda}\,\Theta\left(\frac{R_z}{\lambda}\right)\,
{I}_1\left(\frac{A_z}{\lambda}\right),
\label{eq30}
\end{equation}
\noindent
where
\begin{equation}
\Theta(t)\equiv\frac{16}{t^5}\left[t^2\sinh(t)-3t\cosh(t)
+3\sinh(t)\right].
\label{eq31}
\end{equation}

The constraints on the parameters of Yukawa-type interaction
can be now obtained from the inequality
\begin{equation}
\left|\gamma_z^{\rm Yu}(a_i)\right|\leq\Delta_i\gamma_z,
\label{eq32}
\end{equation}
\noindent
where the left-hand side is defined in Eq.~(\ref{eq30}).
The computational results are presented by the solid line
in Fig.~5. Within the interaction range $\lambda\leq 2\,\mu$m
the best constraints given by the solid line were obtained in an
equilibrium situation when the temperatures of the plate and
of the environment are equal to 310\,K. In so doing the
magnitude of the absolute error
$\Delta_1\gamma_z=3.06\times 10^{-5}$ was used which occurs at the
separation distance $a_1=6.88\,\mu$m \cite{39}.
 Within the interaction range $\lambda\geq 5\,\mu$m
the best constraints are also obtained in an equilibrium
situation. Here, the measurement at the separation
$a_2=9.95\,\mu$m was used with the respective absolute error
equal to $\Delta_2\gamma_z=1.41\times 10^{-5}$ \cite{39}.
In an intermediate interaction range of $\lambda$ from 2 to
$5\,\mu$m the best constraints follow from out of equilibrium
measurement data, where the temperature of an environment was the
same as before, but the temperature of the plate was equal to
479\,K. The best constraints were found from the measurement
performed at $a_3=7.44\,\mu$m with
$\Delta_3\gamma_z=2.35\times 10^{-5}$ \cite{39}.

For comparison purposes, the long-dashed line in Fig.~5 shows
the constraints obtained in Ref.~\cite{24} from the torsion
pendulum experiment of Ref.~\cite{23}. It is seen that constraints
 obtained from the torsion pendulum experiment are from about one
 to two orders of magnitude stronger (depending on the interaction
range) than the constraints given by the solid line. It is important
to bear in mind, however, that the constraints from the
measurement of the Casimir-Polder force (the solid line) were
determined at a 67\% confidence level, whereas the confidence
level of constraints given by the long-dashed line cannot be
determined. This is caused by some uncertain features of the
experiment \cite{23} discussed in Refs.~\cite{21,22}.
The short-dashed line in Fig.~5 presents the strongest
constraints in this interaction range following from
gravitational experiments. These are high confidence
constraints obtained in Ref.~\cite{17}. A new experimental
technique suited for obtaining
stronger constraints on the Yukawa-type interaction from the
measurements of the Casimir-Polder force is
discussed in Ref.~\cite{56}.

Recently one more determination of the limits on the
Yukawa-type interaction in the range of about $1\,\mu$m from
the measurement of the Casimir force using a torsion
balance has been reported \cite{57}. In this experiment a glass
plate and a spherical lens of $R=20.7\,$cm radius were used
as the test bodies, both coated with a 20\,nm Cr layer and
then with a $1\,\mu$m layer of Au. This experiment is not
an independent measurement of the Casimir force, like
experiments in Refs.~\cite{26,27,28,34,35,36}, because
the measurement data for the gradient of the force were
fitted to the sum of two functions: the gradient of the
expected electrostatic force and the gradient of the Casimir
force taking into account the conductivity and
roughness effects. To obtain the limits on $\alpha$,
the fit to the sum of three functions was performed with
inclusion of the Yukawa-type force and $\lambda$ as an
additional fitting parameter. The Yukawa-type force was
calculated using the results of Ref.~\cite{24}. The obtained
limits determined at a 95\% confidence level are shown
as the grey line in Fig.~5. As can be seen in Fig.~5,
these limits are a bit stronger than the limits obtained
from the measurement of the Casimir-Polder force (the
solid line), but much weaker than the limits shown by the
long-dashed line which are obtained \cite{24} from the
measurement data of Ref.~\cite{23}.  The authors of
Ref.~\cite{57} note, however, that the accuracy of the
experiment \cite{23} was overestimated and compare their
limits not with the results of Ref.~\cite{24}, but with the
results of Ref.~\cite{25} where much weaker constraints from
the same measurement data \cite{23} were found. {}From this they
conclude that their limits are stronger than those obtained
on the basis of Ref.~\cite{23}.
It seems that this conclusion is not well
justified. The point is that Ref.~\cite{24} had already
taken into account that the accuracy of the
experiment \cite{23} was overestimated (this is discussed
in \cite{24} in detail). The difference between the constraints
obtained in Refs.~\cite{24} and \cite{25} using the same
data of Ref.~\cite{23} is explained by the fact that in
\cite{24} the corrections due to the surface roughness,
finite conductivity of the boundary metal and nonzero
temperature were taken into account, whereas in \cite{25}
they were disregarded. Thus, in fact the constraints of
Ref.~\cite{57} are weaker than those following from the
measurements of Ref.~\cite{23}. The advantage of the constraints
of Ref.~\cite{57} is that they are obtained at a high confidence
level, whereas the confidence level of the constraints obtained
from the data of Ref.~\cite{23} cannot be determined on a
solid basis.

\section{Conclusions and discussion}

In the foregoing we have considered constraints on the
Yukawa-type corrections to Newtonian gravity in micrometer
and submicrometer interaction ranges following from the
measurements of the Casimir and Casimir-Polder forces.
This field of research has already received wide recognition
as an important adjunct to the gravitational experiments of
E\"{o}tvos- and Cavendish-type. Novel constraints on the
parameters of Yukawa-type interactions from the Casimir effect
were obtained at a high confidence level and were used for
constraining  masses of predicted light elementary particles
and other parameters of the theory of fundamental interactions
beyond the Standard Model \cite{58}.

In this paper we have analyzed several experiments on measuring
the Casimir and Casimir-Polder force which were not used up to
date for constraining corrections to Newtonian gravity.
The most striking result obtained above is that recent experiment
on measuring the lateral Casimir force between corrugated
surfaces of a sphere and a plate \cite{38} leads to a great
strengthening of the previously known high confidence constraints
up to a factor of $2.4\times 10^7$. The constraints obtained
from this experiment are shown to be the strongest in the interaction
range from 1.6 to 14\,nm. This raises a question on the role of
corrugated surfaces in other experimental configurations used to
obtain constraints on the parameters of Yukawa-type
interactions.

The influence of sinusoidal corrugations on the Yukawa force was
calculated above by using the approximate method of geometrical
averaging. This method is applicable under several conditions
which are well satisfied for the experimental configurations
considered in the paper. In fact, the geometrical averaging works
better for the Yukawa force than for the Casimir force. The point
is that the Yukawa force is static and is not influenced by the
diffraction-type effects which are essential for the Casimir
force (the better applicability of the geometrical averaging to the
electrostatic than to the Casimir force was also noted in
Ref.~\cite{38}). We have shown that the replacement of a plane
plate with a sinusoidally corrugated plate in the experiment on
measuring the normal Casimir force in a sphere-plate
configuration \cite{28} would strengthen the constraints
obtained from the measurement data of Ref.~\cite{28}  in
Ref.~\cite{36} by a factor of $1.1\times 10^{12}$.

Next, we have obtained constraints on the Yukawa-type interaction
from the measurement data of the dynamic experiment using a
micromechanical torsional oscillator \cite{36} when the plane
plate is replaced with the
corrugated one.
It was shown that up to an order of magnitude strengthening
 is achievable in this way in comparison with
the previously obtained constraints.

We have also obtained constraints from the measurement data of
two dynamic experiments which were not used previously for
constraining the Yukawa-type interaction. These are the
measurement of the gradient of the Casimir pressure in the
configuration of two parallel plates \cite{40} and of the
gradient of the thermal Casimir-Polder force between an atom and a
dielectric plate \cite{39}. For the parallel plate experiment,
the obtained constraints are weaker than those obtained from
other experiments, but they can be significantly strengthened
by replacing a Cr metal coating with a thicker Au coating.
For the experiment dealing with the Casimir-Polder interaction,
a frequency shift of the condensate oscillating due to the
presence of Yukawa-type interaction was found. It was shown
that the obtained constraints are weaker than those found in
Ref.~\cite{24} from the torsion pendulum experiment \cite{23}.
It was also shown that the constraints following from the
measurement of the Casimir-Polder interaction are of almost
the same strength as the constraints obtained in recent
experiment \cite{57}. For both the experiment with two parallel
plates and the measurement of the Casimir-Polder force, the
replacement of a plane plate with a grating does not lead to a
notable improvement in the strength of constraints because of
the relatively large separation distances between the test bodies
in these experiments.

Thus, we have shown that the use of the measurement data of recent
experiment on the lateral Casimir force leads to more than seven
orders of magnitude strengthening in the previously obtained
constraints. We have demonstrated that even higher promise is expected
from the measurement of the normal Casimir force by means of an
AFM if the plane plate were replaced with a corrugated plate.
These results and the analysis of some other experiments
confirm prospective future trends for obtaining stronger
constraints on non-Newtonian gravity from
the measurement of both the Casimir and Casimir-Polder
forces.

\section*{Acknowledgments}

The authors thank FAPES-ES/CNPq (PRONEX) for partial financial support.
G.L.K.\ and V.M.M.\ are grateful to the Federal University of
Para\'{\i}ba (Jo\~{a}o Pessoa, Brazil), where this work was
performed, for kind hospitality.
V.B.B.\ and C.R.\ also thank CNPq  for partial financial support.


\begin{figure}[b]
\vspace*{-11cm}
\centerline{\hspace*{7cm}
\includegraphics{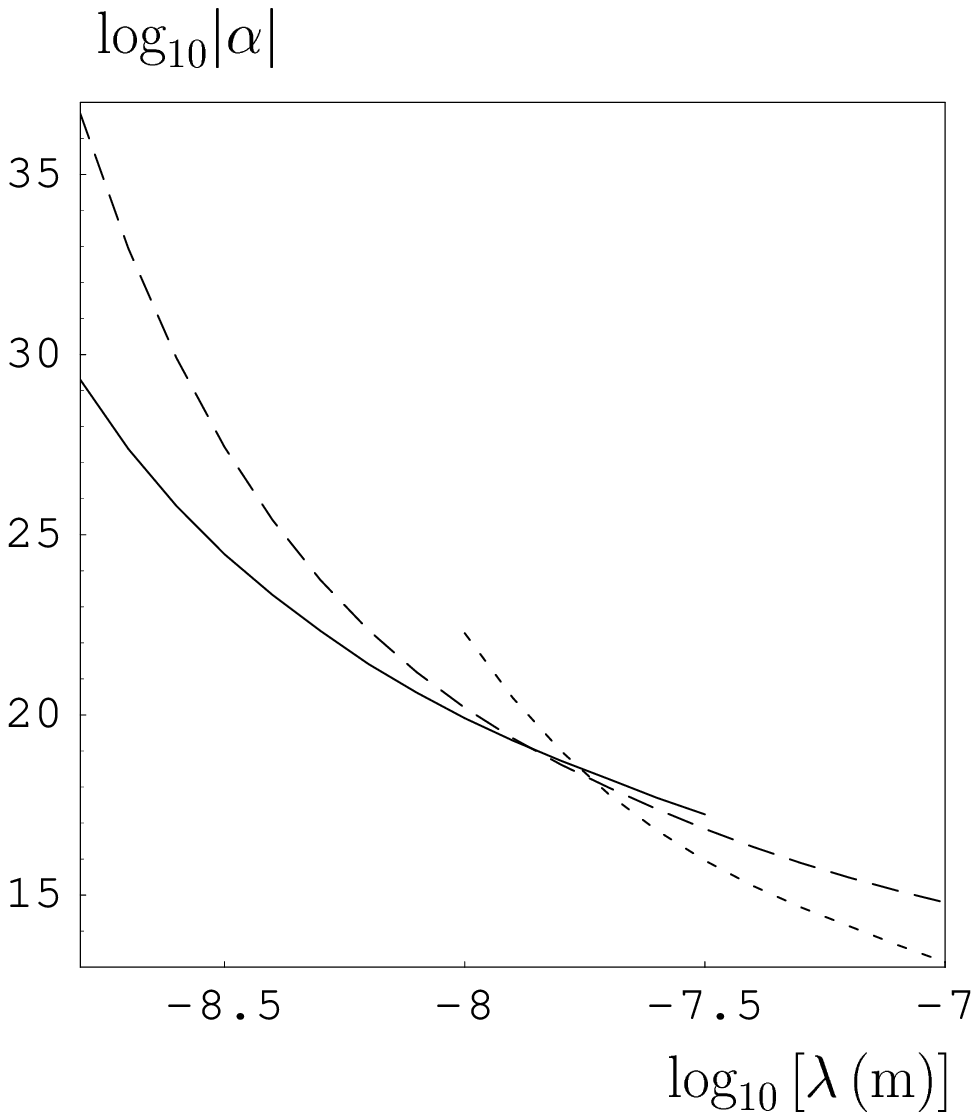}
}
\vspace*{-6cm}
\caption{Constraints on the parameters of Yukawa-type interaction
from measurements of the lateral Casimir force between corrugated
surfaces (the solid line), and from measurements of the
normal Casimir force by means of an atomic force microscope
(the long-dashed line), and a micromachined oscillator
(the short-dashed line).
The allowed regions in the $(\lambda,\,\alpha)$-plane lie
beneath the lines.
}
\end{figure}
\begin{figure}[b]
\vspace*{-11cm}
\centerline{\hspace*{7cm}
\includegraphics{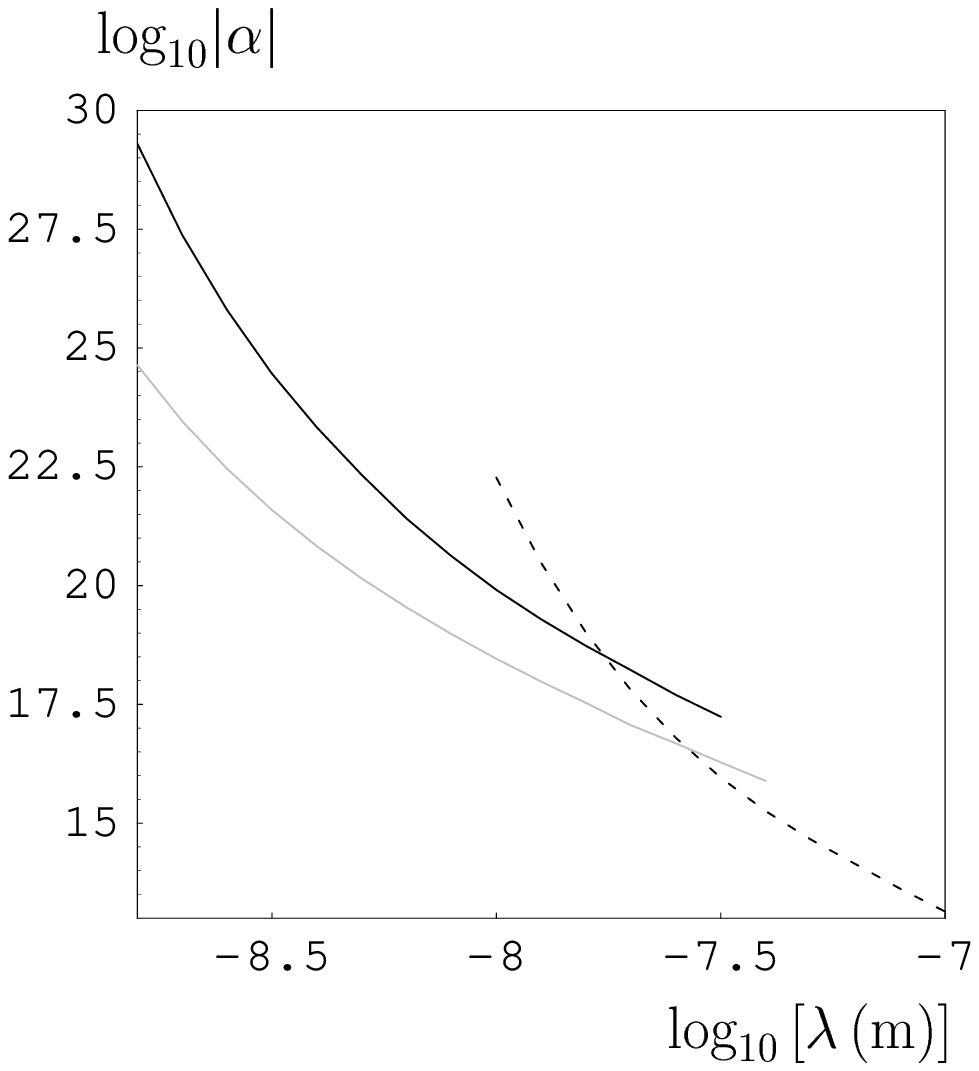}
}
\vspace*{-6cm}
\caption{Constraints on the parameters of Yukawa-type interaction
from proposed  measurements of the normal Casimir force between
a smooth sphere and a corrugated plate (the grey line),
from measurements of the lateral Casimir force between corrugated
surfaces (the solid line), and from measurements of the
normal Casimir force using a micromachined oscillator
(the short-dashed line).
The allowed regions in the $(\lambda,\,\alpha)$-plane lie
beneath the lines.
}
\end{figure}
\begin{figure}[b]
\vspace*{-11cm}
\centerline{\hspace*{7cm}
\includegraphics{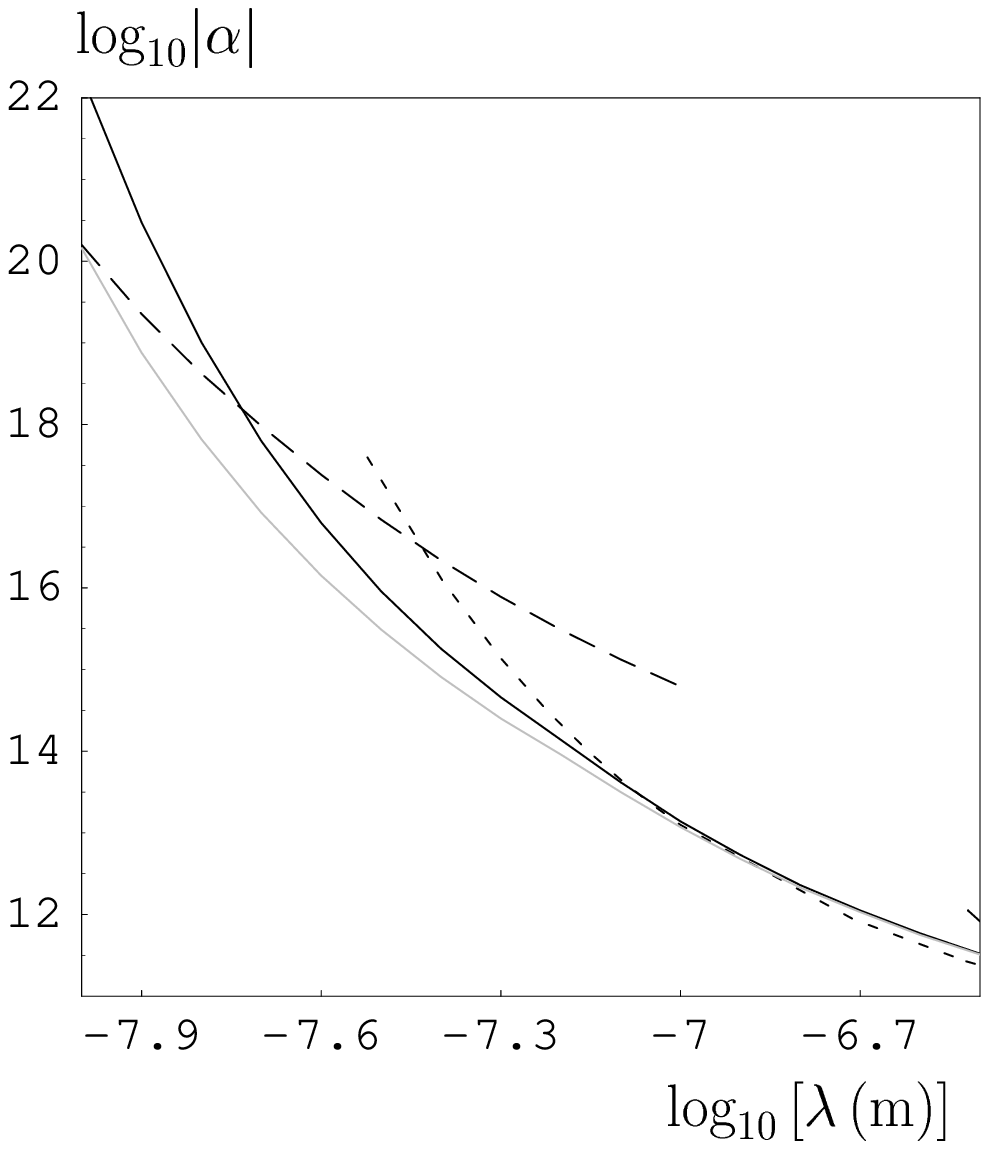}
}
\vspace*{-6cm}
\caption{Constraints on the parameters of Yukawa-type interaction
from the experiment  using a micromachined oscillator
(the solid line), from the same experiment where the plane plate
is replaced with the corrugated plate (the grey line),
from measurements of the
normal Casimir force by means of an atomic force microscope
(the long-dashed line), and from the Casimir-less experiment
(the short-dashed line).
The allowed regions in the $(\lambda,\,\alpha)$-plane lie
beneath the lines.
}
\end{figure}
\begin{figure}[b]
\vspace*{-11cm}
\centerline{\hspace*{7cm}
\includegraphics{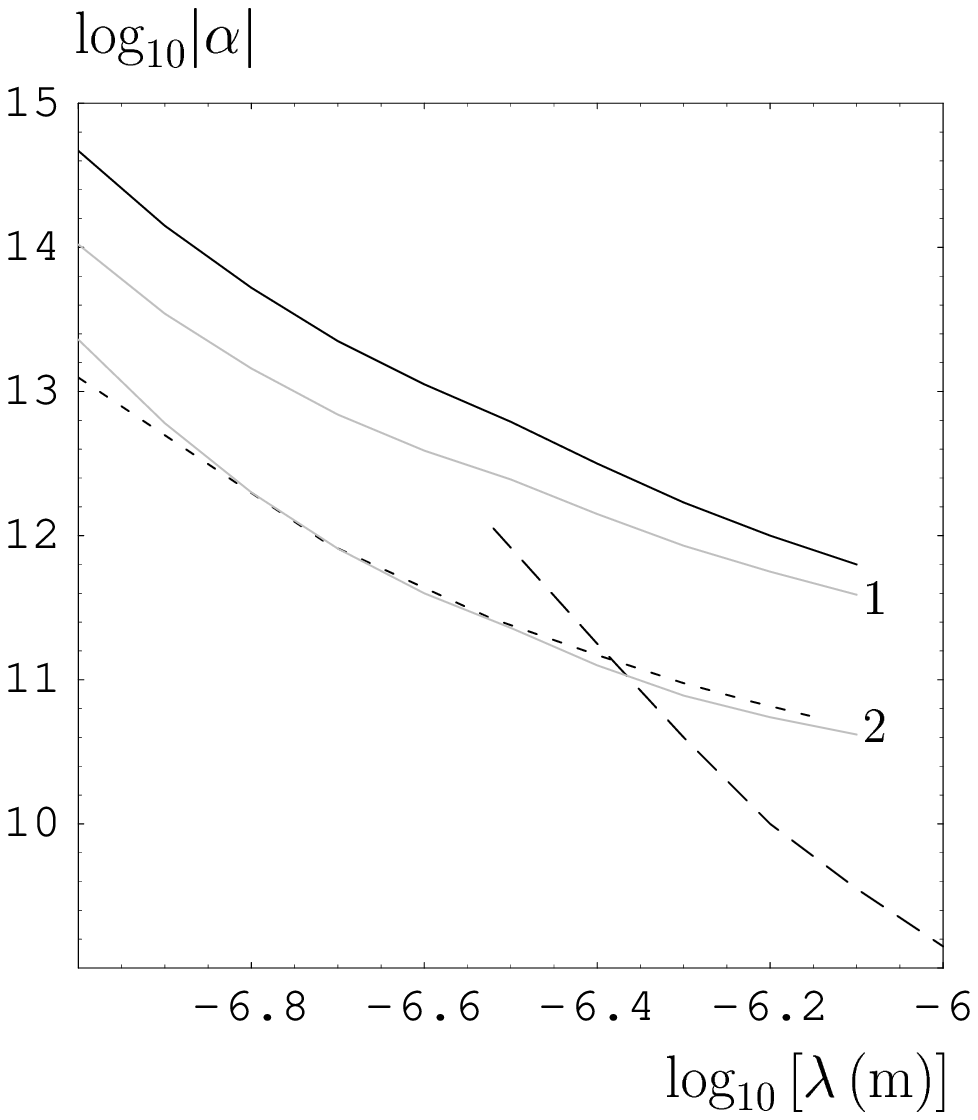}
}
\vspace*{-6cm}
\caption{Constraints on the parameters of Yukawa-type interaction
from the experiment with two parallel plates (the solid line),
from the same experiment with a Cr coating replaced with an Au
coating of the same thickness (the grey line 1),
from the same experiment with ten times thicker Au
coating (the grey line 2),
from the Casimir-less experiment (the short-dashed line),
and from the measurement of the Casimir force by means of a torsion
pendulum (the long-dashed line).
The allowed regions in the $(\lambda,\,\alpha)$-plane lie
beneath the lines.
}
\end{figure}
\begin{figure}[b]
\vspace*{-11cm}
\centerline{\hspace*{7cm}
\includegraphics{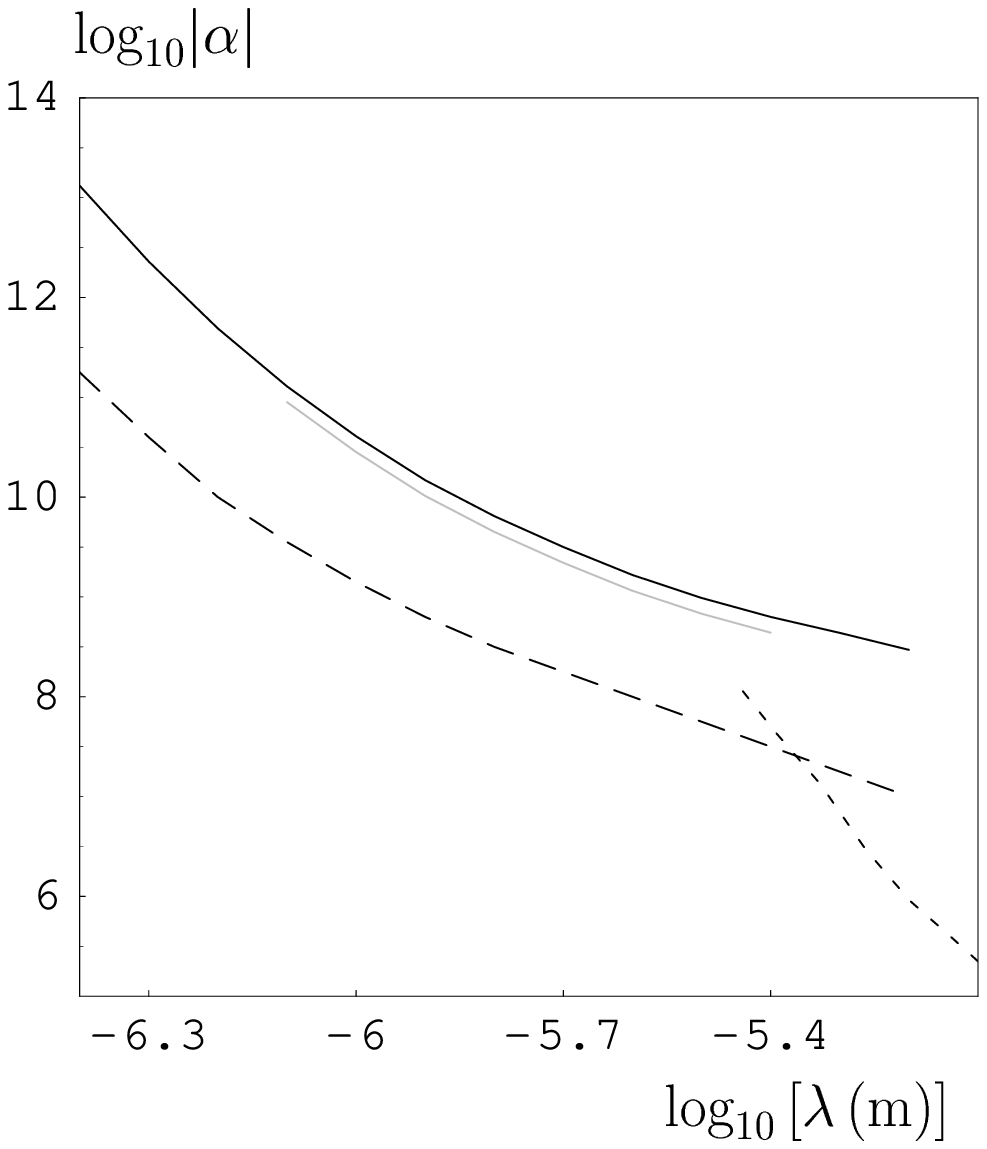}
}
\vspace*{-6cm}
\caption{Constraints on the parameters of Yukawa-type interaction
from measurements of the Casimir-Polder force (the solid line),
from the experiment using a torsion balance (the grey line),
from the measurement of the Casimir force by means of a torsion
pendulum (the long-dashed line), and from the gravitational
experiment of Ref.~\cite{17} (the short-dashed line).
The allowed regions in the $(\lambda,\,\alpha)$-plane lie
beneath the lines.
}
\end{figure}
\end{document}